\documentclass[journal]{IEEEtai}

\usepackage[colorlinks,urlcolor=blue,linkcolor=blue,citecolor=blue]{hyperref}
\usepackage{cite}
\usepackage{color,array}
\usepackage{graphicx}
\usepackage{amsmath}
\usepackage{amssymb}
\usepackage{caption}
\usepackage[numbers]{natbib}
\usepackage{mathtools} 
\usepackage{xcolor}
\usepackage{soul}
\usepackage{fancyhdr}
\fancypagestyle{zenodo}{\fancyhf{} \fancyfoot[L]{*\url{https://zenodo.org/record/7301863}}}

\fancypagestyle{github}{\fancyhf{} \fancyfoot[L]{**\url{https://github.com/mahindrautela/DeepGenerator_CompositeMaterialGWrepresentations}}}

\begin{document}

\title{Towards deep generation of guided wave representations for composite materials} 

\author{{Mahindra Rautela, J. Senthilnath, \IEEEmembership{Senior~Member,~IEEE}, Armin Huber and S. Gopalakrishnan}
\thanks{J.S. acknowledge funding from the Accelerated Materials Development for Manufacturing Program at A*STAR via the AME Programmatic Fund by the Agency for Science, Technology and Research under Grant No. A1898b0043}
\thanks{Mahindra Rautela is with the Department of Aerospace Engineering, Indian Institute of Science, Bangalore, India (e-mail: mrautela@iisc.ac.in).}
\thanks{J. Senthilnath is with the Institute for Infocomm Research, A*STAR, Singapore, 138632 (e-mail: J\_Senthilnath@i2r.a$-$star.edu.sg).}
\thanks{Armin Huber is with Center for Lightweight Production Technology, German Aerospace Center (DLR), Augsburg, Germany}
\thanks{S. Gopalakrishnan is with the Department of Aerospace Engineering, Indian Institute of Science, Bangalore, India (e-mail: krishnan@iisc.ac.in).}}

%\markboth{Journal of IEEE Transactions on Artificial Intelligence, Vol. 00, No. 0, Month 2020}
%{Mahindra Rautela \MakeLowercase{\textit{et al.}}: Bare Demo of IEEEtai.cls for IEEE Journals of IEEE Transactions on Artificial Intelligence}

\maketitle
\begin{abstract}
Laminated composite materials are widely used in most fields of engineering. Wave propagation analysis plays an essential role in understanding the short-duration transient response of composite structures. The forward physics-based models are utilized to map from elastic properties space to wave propagation behavior in a laminated composite material. Due to the high-frequency, multi-modal, and dispersive nature of the guided waves, the physics-based simulations are computationally demanding. It makes property prediction, generation, and material design problems more challenging. In this work, a forward physics-based simulator such as the stiffness matrix method is utilized to collect group velocities of guided waves for a set of composite materials. A variational autoencoder (VAE)-based deep generative model is proposed for the generation of new and realistic polar group velocity representations. It is observed that the deep generator is able to reconstruct unseen representations with very low mean square reconstruction error. Global Monte Carlo and directional equally-spaced samplers are used to sample the continuous, complete and organized low-dimensional latent space of VAE. The sampled point is fed into the trained decoder to generate new polar representations. The network has shown exceptional generation capabilities. It is also seen that the latent space forms a conceptual space where different directions and regions show inherent patterns related to the generated representations and their corresponding material properties.
\end{abstract}

\begin{IEEEImpStatement}
AI-accelerated property prediction, discovery, and design of materials have emerged as a new research front with many promising features. There are many investigations on different materials, but no emphasis is placed on composite materials. Among many challenges, the unavailability of datasets for composite materials is a significant roadblock. This is because conducting multiple experiments is costly and cumbersome, and performing simulations is time-taking and demands computational resources. In order to accelerate and scale the prediction, discovery, and design, a deep generation approach is proposed for composite materials. The current research requires limited physical simulations to train a deep generator network. The generator can generate enormous data, eliminating the demerits of both experiments and simulations. The work is novel in terms of the deep generation approach as well as the applications for composite materials.
\end{IEEEImpStatement}


\section{Introduction}
Fiber-reinforced-polymer (FRP) composites play a dominant role in aerospace and related industries due to their high specific strength. With improvements in composite manufacturing, their use for low and high technology applications is increasing. Guided wave propagation is a high-frequency, multi-modal, and dispersive phenomenon, which is studied in the literature for different materials and structures \cite{rose2014ultrasonic}. Study of wave propagation finds applications in a variety of problems, some of them are damage identification \cite{rautela2022delamination}, material property prediction \cite{rautela2021inverse}, geometrical parameters estimation \cite{ponschab2019simulation}, and impact force mitigation \cite{raut2021elastic}. 

Compared with experimental measurements, computational simulation requires less time and is advantageous in studying wave propagation behavior for multiple test scenarios. Apart from this, simulations give complete control of the variables and physics of the problem \cite{iwasaki2019machine}. In most of the simulations related to wave propagation, a physics-based model maps system properties (material and geometry) and external fields (e.g., forces, moments) to wave propagation behavior (displacement, velocity, acceleration, strain, and stress fields). Despite having multiple simulation methods, wave propagation analysis in composite structures is complicated, time-taking, and limited by modeling assumptions \cite{willberg2015simulation}. In such cases, inverse property prediction, generation, and design become more challenging. Some of the difficulties are multiple local optima, high-dimensionality, and expensive computations \cite{wang2020deep}. Thus, it is necessary to develop new methods for accelerating such problems for the space of other composite materials \cite{liu2017materials}. 
In other fields of material science, there are substantial advances in machine learning and deep learning for the prediction, discovery, and design of materials. These three problems are hierarchically described in detail in Ref.~\cite{cai2020machine}. The property prediction problem aims to inversely find the material property from the characteristics/behavior of the material \cite{rautela2021inverse}. The discovery problem is focused on generation to find a new set of material properties or characteristics/behavior which are not present in the dataset. In the literature, predictive models and generative models are used for the discovery of oxides \cite{huang2017material}, nanomaterials and alloys \cite{li2018deep}, heat conduction materials \cite{guo2018indirect}, thermoelectrics \cite{iwasaki2019machine}, metamaterials \cite{wang2020deep}, inorganic solid materials \cite{noh2020machine}, and nanoparticles \cite{mekki2021two}. The material design problem is an optimization problem that needs an objective function, a possible set of constraints, and an algorithm to search the latent space to find the optimal class of materials \cite{wang2020deep}. However, for composite materials, there are very few investigations towards property prediction problems and no studies on the generation and design problems. In our previous works, for the first time, we have successfully implemented deep learning strategies to inversely predict the material properties of composites using guided waves \cite{rautela2021inverse}. The technique is seen to have leverage over traditionally used optimization schemes like genetic algorithms \cite{vishnuvardhan2007genetic}, and simulated annealing \cite{cui2019identification}. One of the difficulties of the existing methodologies is the expensive forward physics-based wave propagation simulations. In order to scale up and accelerate the prediction problem to more challenging generation and design problems, running such computationally demanding simulations create a bottleneck. Deep generative models can replace these expensive simulators, which necessarily have inherent capabilities for accelerating generation and design problems for the space of other composite materials \cite{liu2017materials}.

In this paper, the stiffness matrix method is revisited for generating polar group velocity representation corresponding to the space of different composite materials. A variational autoencoder (VAE) based deep generative model is proposed to map polar group velocities representations to a lower dimensional latent space. The continuous and complete latent space of VAE guarantees the generative property. A Monte-Carlo sampler is used to sample the latent space, and a decoder maps the latent point to generate new and realistic samples. The study of directional variations in the latent space is used to understand the inherent patterns hidden in the latent space. The expensive physics-based simulations can be replaced with a prediction model and a deep generator for design, optimization and related problems. The paper is organized as follows: Section-\ref{sec:theory} contains the background of variational autoencoders and physical modeling. Section-\ref{sec:results} presents results and discussion on training-testing and generation of guided wave representations. The paper is concluded in Section-\ref{sec:conclusion}.

%  Begin NEW SECTION
\section{Theoretical Background} \label{sec:theory}
\subsection{Deep generation via variational autoencoding}\label{ssec:theory}
Deep generative models learn complex high-dimensional probabilistic governing distribution that can generate new samples. Deep generative models can be cost function based like generative adversarial networks (GAN) \cite{goodfellow2014generative} and variational autoencoders \cite{kingma2013auto}, or energy-based like deep belief networks (DBN) \cite{hinton2006fast} and deep boltzmann machines (DBM) \cite{hinton2012better}. VAE and GAN are the most popular for generative modeling. A GAN is trained to approach a Nash equilibrium of the two-player zero-sum min-max game. In GAN, a generator network competes against a discriminator network to generate new samples. In contrast, VAE trains an encoder and decoder (generator) simultaneously, which explicitly learns likelihood distribution. Although GAN is considered a better generator but VAE provides a continuous and complete latent space, which is generally less explicit from GANs \cite{goodfellow2016deep}. Also, GAN faces mode collapse and stability issues for some problems \cite{srivastava2017veegan}. Since our aim is to utilize the latent space representations, VAE is more suitable for the study.

\begin{figure*}[h!]
	\centering
	\captionsetup{justification=centering}
	\includegraphics[width=0.7\textwidth]{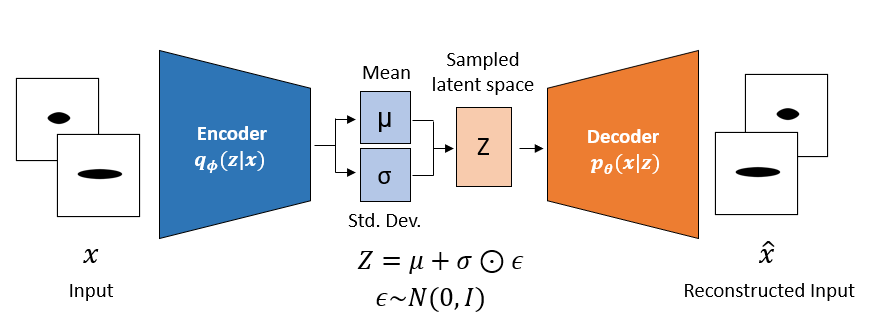}
	\caption{Dual input architecture of a variational autoencoder.}
	\label{fig:vae}
	\vspace{2mm}
\end{figure*}

VAE aims to model a generative process using a joint probability distribution $p_{\theta}(x,z)$ over observed data ($x$) and latent space ($z$). Using Bayes' rule,  $p_{\theta}(x,z)=p_{\theta}(x|z)p_{\theta}(z)$, where $\theta$ are the model parameters, $p_{\theta}(x|z)$ is the likelihood and $p_{\theta}(z)$ is the prior distribution. The likelihood can be modeled as a deep neural network (decoder) parameterized by $\theta$. However, even with simple prior distribution on latent variables like $\mathcal{N}(0,I)$, the marginal distribution $p_{\theta}(x)$ used to calculate likelihood is intractable. VAE introduces another deep neural network $q_{\phi}(z|x)$ (encoder) parameterized by $\phi$ to tackle intractability. It is an inference model which maps the data $x$ to the latent space $z$ by approximating the posterior distribution $p_{\theta}(z|x)$. Fig.~\ref{fig:vae} shows a pictorial representation of a dual channel input network architecture of a VAE. The likelihood function for training the encoder-decoder network is approximated by the evidence lower bound (ELBO). 
 
\begin{equation}
	\text{ELBO} = \mathbb{E}_{q_{\phi}(z|x)} [\log p_{\theta} (x|z)] - \mathbb{E}_{q_{\phi}(z|x)} \Big[\log \frac{q_{\phi}(z|x)}{p_{\theta} (z|x)}\Big]
\end{equation}

The first term in the equation can be viewed as the reconstruction error between the original and reconstructed data due to information loss in the dimensionality reduction process. The second term is the Kullback–Leibler (KL) divergence $D_{KL}[q_{\phi}(z|x)||p_{\theta}(z|x)]$, which is a measure of similarity (or divergence) between a posterior distribution and the approximate distribution. A normal autoencoder is trained only with reconstruction loss. The latent space of a normal autoencoder is neither continuous nor complete, which restricts its generation ability \cite{rautela2022delamination}.

VAE assumes standard normal prior distribution $p_{\theta}(z) \sim \mathcal{N}(0,I)$ and a Gaussian distribution for posterior $p_{\theta}(z|x) \sim \mathcal{N}(\mu,\sigma)$. With the reparameterization trick, the usual Monte-Carlo estimator for ELBO used in stochastic gradient descent (SGD) is reduced to the following optimization problem Eq.~\ref{eq:elbo2} \cite{wang2020deep}.

\begin{equation}\label{eq:elbo2}
\begin{aligned}[b]
	&[\theta,\phi] = \underset{\theta,\phi}{\mathrm{argmin}} \Big(-\Big[\frac{1}{N}\sum_{n}^{N} \log(p_{\theta}(x|z))- \\
	& \frac{1}{2}\sum_{l}^{L}(1+\log(\sigma^2_l)-\sigma^2_l-\mu^2_l)\Big]\Big)
\end{aligned}
\end{equation}
where $N$ and $L$ are the number of samples and dimensions of latent space, $\theta$ and $\phi$ are the parameters of encoder and decoder, respectively.

\subsection{Physical modeling of guided wave propagation} \label{ssec:dataset}
Computational simulations and experiments are two conventional methods used in the field of materials science. It is time-taking, costly, and cumbersome to produce multiple composite specimens with different material properties and perform experiments. Compared to experiments, computational simulation requires less time along with complete control over the relevant variables. In the material science community, the datasets for prediction, discovery, and design problems are majorly created via Density Functional Theory (DFT) based quantum-mechanical atomistic simulations \cite{curtarolo2013high,Zekun2022MatSci}. Due to the lack of availability of a dataset for our problem, we have created the dataset by performing physics-based macroscopic modeling and simulations. For this, the stiffness matrix method (SMM) is used for producing high-fidelity guided wave simulations for different composite materials.

Ultrasonic Lamb waves are the superposition of bulk waves in waveguides. Due to the presence of boundaries and interfaces, phase velocity and group velocity of the wave becomes frequency-dependent (dispersion) \cite{rose2014ultrasonic}. Along with this, the velocities also depend on the propagation direction for anisotropic composites. In acoustic field theory, the Christoffel equation is a combination of three equations. These equations are strain-displacement relation, equation of motion and elastic constitutive equation, combining them gives Eq.~(\ref{eq.07}) \cite{huber2018classification}.
\begin{equation}\label{eq.07}
	\rho\frac{\partial^2 u_{i}}{\partial t^2}=c_{ijkl}\frac{\partial^2 u_l}{\partial x_j\partial x_k}.
\end{equation}
where $\rho$ is the density of the solid, $u$ is the displacement field and $c_{ijkl}$ is the stiffness tensor. To solve the above equation, $u_i$ can be written in terms of bulk wave amplitudes $U_i$ as shown in Eq.~(\ref{eq.08}).
\begin{equation}\label{eq.08}
	(u_1,u_2,u_3)=(U_1,U_2,U_3)\mathrm e^{\mathrm i\xi(x_1+\alpha x_3-c_\mathrm pt)},
\end{equation}

In the above equation, $c_\mathrm p$ is the phase velocity component along $x_1$, $\xi$ ($=\omega/c_\mathrm p$) is the wavenumber along $x_1$, and $\alpha$ is the ratio of the wavenumber of bulk waves along the $x_3$ and $x_1$-directions ($\alpha=\zeta_3/\zeta_1=\zeta_3/\xi$). Eq.~(\ref{eq.08}) can be substituted into the Eq.~(\ref{eq.07}) to obtain the Christoffel matrix 

\begin{align*}\label{eq.09}
    \scriptsize
    \begin{bmatrix}
		C_{11}-\rho c_\mathrm p^2+C_{55}\alpha^2&C_{16}+C_{45}\alpha^2 &(C_{13}+C_{55})\alpha\\
		&C_{66}-\rho c_\mathrm p^2+C_{44}\alpha^2&(C_{36}+C_{45})\alpha\\
		\mathrm{sym}&&C_{55}-\rho c_\mathrm p^2+C_{33}\alpha^2 \\
	\end{bmatrix} \\
	\scriptsize
	\begin{pmatrix}
		U_1 \\
		U_2 \\
		U_3 
	\end{pmatrix}=0
\end{align*}

\begin{figure*}
	\centering
	\includegraphics[width=0.8\textwidth]{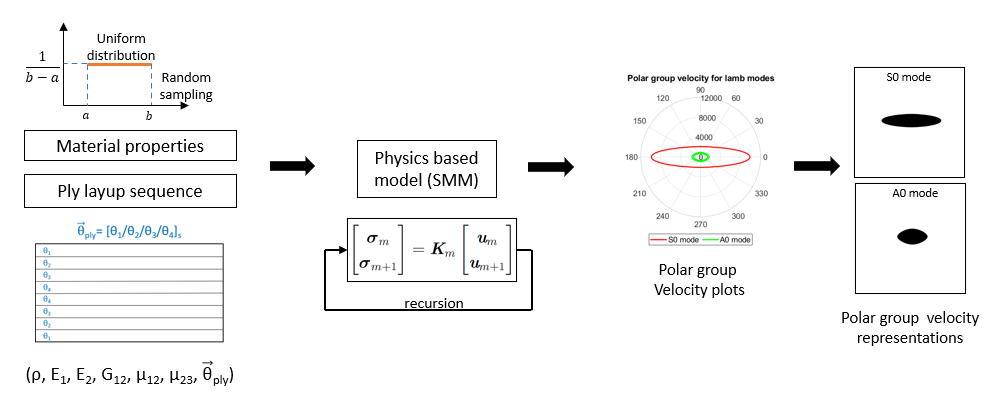}
	\caption{Detailed schematic of forward physics-based model: Material properties ($\rho$,$E_1$,$E_2$,$G_{12}$,$\nu_{12}$,$\nu_{23}$) and ply-layup sequence (constant) are the inputs to the model and polar representations are the outputs \cite{rautela2021inverse}}
	\label{fig:forward}
\end{figure*}

The matrix contains stiffness matrix coefficients $C_{ij}$, wavenumber ratio $\alpha$, phase velocity $c_p$ and density $\rho$. In order to obtain non-trivial solutions for $U_1$, $U_2$, and $U_3$, the determinant of the matrix vanishes. This yields a sixth-degree polynomial equation, which needs to be solved for $\alpha$. SMM is used extensively in the popular Dispersion Calculator to simulate wave propagation behavior in materials \cite{huber2018classification}. The detailed mathematical formulation of SMM is presented in ~\cite{rautela2021inverse}. The dataset collection process using SMM is pictorially represented in Fig.~\ref{fig:forward}.

For dataset collection, first, we have generated two material sets (Matset-1 and 2) containing six material properties of different transversely isotropic carbon fiber-reinforced polymer (CFRP) composites. The properties are density ($\rho$), Young's modulus in 1 and 2 directions ($E_1$, $E_2$), shear modulus in 1-2 plane ($G_{12}$), Poisson's ratio in 1-2 and 2-3 plane ($\nu_{12}$, $\nu_{23}$). Matset-1 includes material properties of commercial CFRP materials (Table-\ref{tab:ComMat}). Based on the range of these materials, new material properties (Matset-2) are generated using random sampling of a uniform distribution with extended bounds. These bounds are [1304 1760] kg/$\mathrm{m}^3$ for $\rho$, [115 184] GPa for $E_1$, [6 14] GPa for $E_2$, [3 9] GPa for $G_{12}$, [0.2 0.52] for $\nu_{12}$ and [0.23 0.59]) for $\nu_{23}$.

\begin{table*}[ht!]
	\centering
	\captionsetup{justification=centering}
	\setlength{\belowcaptionskip}{0pt}
	\caption{\small Material properties of commercially available CFRP composite materials}
	\addtolength{\tabcolsep}{-1pt} 
	\begin{tabular}{l|c c c c c c}
		\hline
		Commerical CFRP & $\rho$  & $E_1$ & $E_2$ & $G_{12}$ & $\nu_{12}$ & $\nu_{23}$\\
		composite materials &(kg/$\mathrm{m}^3$) & (GPa) & (GPa) & (GPa)& &\\
		\cline{1-7}
		AS4M3502 \cite{SAUSE2018291} & 1550 & 144.6 & 9.6 & 6 & 0.3 & 0.28 \\
		GraphiteEpoxy\_Rokhlin\_2011 \cite{rokhlin2011physical} & 1610 & 150.95 & 12.8 & 8 & 0.47 & 0.45 \\
		SAERTEX7006919RIMR135 \cite{huber2018dispersion} & 1454 & 119.9 & 7.25 & 6 & 0.32 & 0.45 \\
		SigrafilCE125023039 \cite{SAUSE2018291} & 1500 & 128.6 & 6.87 & 6.1 & 0.33 & 0.37 \\
		T300M914 \cite{moll2019open} & 1560 & 139.92 & 10.05 & 5.7 & 0.31 & 0.48\\
		T700M21 \cite{Simon1997}  & 1571 & 125.5 & 8.7 & 4.1 & 0.37 & 0.45\\
		T700PPS \cite{SAUSE2018291} & 1600 & 149.96 & 9.98 & 4.5 & 0.29 & 0.37 \\
		T800M913 \cite{SAUSE2018291} & 1550 & 152.14 & 6.64 & 20 & 0.25 & 0.54 \\
		T800M924 \cite{percival1997study} & 1500 & 161 & 9.25  & 6 & 0.34 & 0.41\\
		T800\_Michel \cite{yu2017feature} & 1510 & 178.96 & 9.17 & 5.5 & 0.36 & 0.53\\
		\hline
	\end{tabular}
	\label{tab:ComMat}
\end{table*}

Both the material property sets are fed into the SMM-based simulator (See Fig.~\ref{fig:forward}). SMM provides polar group velocities at different propagation angles ranging from 0$^{\circ}$ to 360$^{\circ}$. The excitation frequencies range from 20 kHz to 200 kHz in increments of 20 kHz. This is because the majority of Lamb wave-based experiments are performed in this range. Also, it helps in eliminating high-frequency Lamb wave modes, which can complicate the study \cite{mitra2016guided}. Both the fundamental Lamb modes are dispersive in this frequency range. The thickness of the composite is used as 2 mm with 16-layered symmetric laminate.

\begin{figure}[h!]
	\centering
	\includegraphics[width=0.3\textwidth]{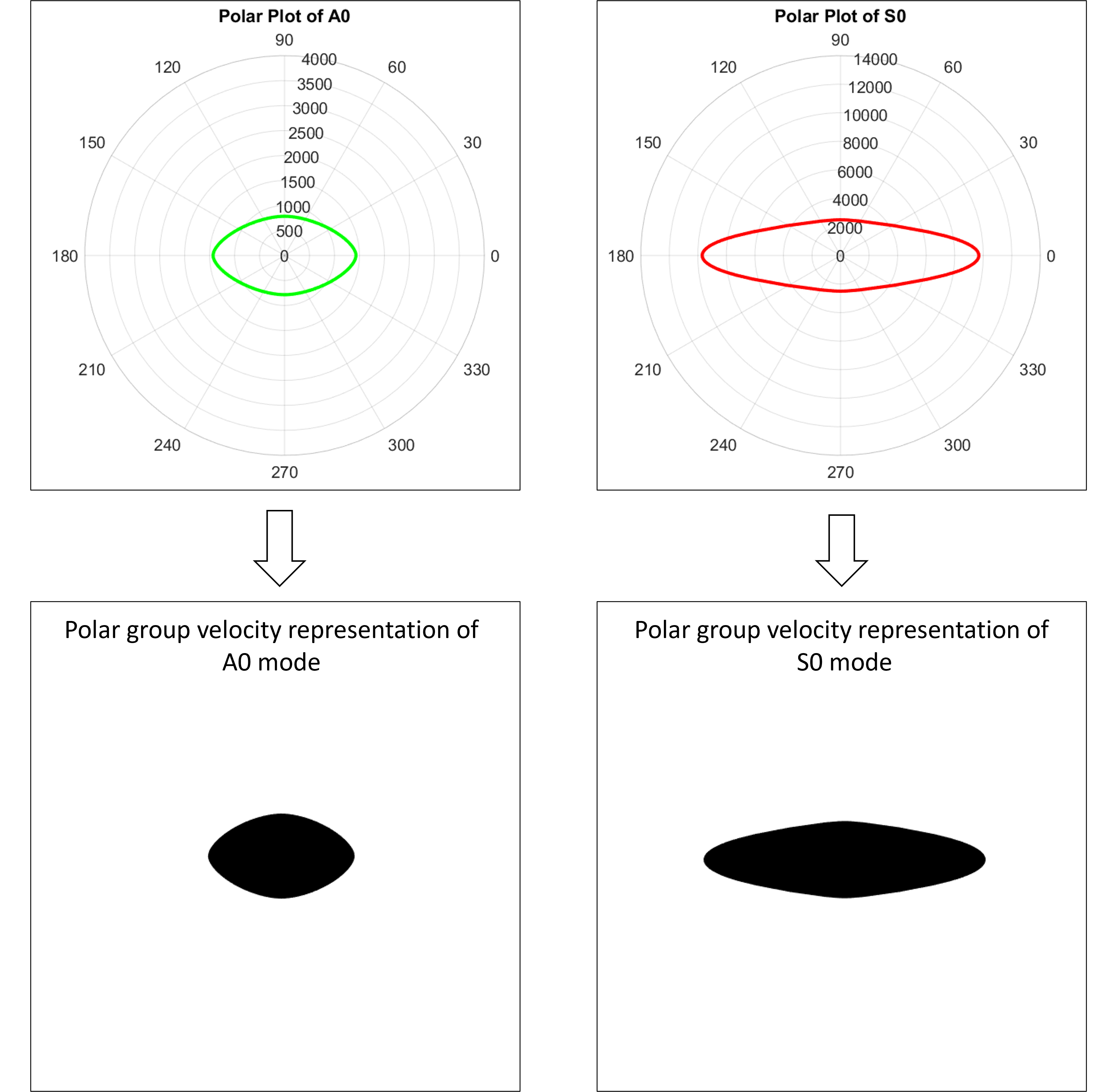}
	\caption{(Top): Polar group velocities plots of $A_0$ and $S_0$ modes for material AS4M3502 (radius = group velocity, circumference = propagation angle), (Bottom): Polar group velocity representations used while training.}
	\label{fig:samples}
\end{figure}

Fig.~\ref{fig:samples} shows SMM calculated polar plots for fiber-reinforced polymer composite AS4M3502 ($\rho$ = 1550 $kg/m^3$, $E_1$ = 144.6 GPa, $E_2$ = 9.6 GPa, $G_{12}$ = 6 GPa, $\nu_{12}$ = 0.3, $\nu_{23}$ = 0.28). In the top plots of Fig.~\ref{fig:samples}, the radial direction is marked with group velocity and circumferential direction with propagation angle. To read the top left plot, the $0^{\circ}$ and $90^{\circ}$ direction wave propagation has a group velocity of nearly 1450 m/s and 750 m/s. To utilize the image processing capabilities of convolutional VAE, the top plots are converted into bottom plots, which are binary-filled images (black and white) called polar group velocity representations. 

Dataset-1 includes polar representations corresponding to 10 materials with 10 different frequencies for each Lamb mode and three different ply-layup sequences, i.e., unidirectional, cross-ply, and quasi-isotropic. On the other hand, Dataset-2 contains representations of 987 materials with 10 different frequencies for each Lamb mode and a unidirectional layup. The number of samples in Dataset-1 and Dataset-2 is 600 samples and 19740, respectively. The dataset is available open source at the weblink* \cite{rautelaGWdataset2022}. The dataset collected from the forward physics-based simulator is used to assist in the generation of realistic representations via VAE.
\thispagestyle{zenodo}
One of the essential points to explain here is the experimental collection of polar plots. In ultrasonic-guided wave experiments, an array of sensors are used to measure the wave response of the structure at a different angle of propagation (sensor's location with respect to excitation). The group velocities can be calculated from raw time-series signals corresponding to the different angles of propagation even in the presence of noise \cite{wang2007group,yue2020scalable,malik2021direct}. It can be directly converted to polar plots (top plots in Fig.~\ref{fig:samples}) and then to polar group velocity representations (bottom plots in Fig.~\ref{fig:samples}). 

% Begin NEW SECTION
\section{Results and Discussion} \label{sec:results}
\subsection{Training and Testing}
The dataset-2 contains 9,870 polar group velocity representations each for $A_0$ and $S_0$ modes, which makes an overall of 19740 representations. The dataset is split randomly in 85:15 way into training and validation sets to make sure the network is not underfitting or overfitting. For training, the $A_0$ and $S_0$ representations are supplied on two different channels of the variational autoencoders (Fig.~\ref{fig:vae}). The encoder and decoder contain five convolutional layers with a constant 2D kernel of size 3. A 5-dimensional latent space is found suitable after tuning the dimension value based on the overall loss. A learning rate of 1e-4 and batch size of 32 is used. The training is performed for 1500 epochs with an ELBO-based cost function and Adam-based optimization scheme. It takes 7 seconds per epoch on a machine with i9-10920X CPU with 32 GB RAM and NVIDIA Geforce RTX 3080 GPU with 10 GB VRAM. The network is able to achieve overall loss (binary cross-entropy reconstruction loss + KL divergence loss) of 43.637. The reconstruction ability of the network is measured through mean squared error (MSE) and mean absolute error (MAE). The network is able to reach MSE of 7.33e-5 and 7.45e-5 as well as MAE of 4.03e-4 and 4.38e-4 for $A_0$ and $S_0$ modes, respectively.

\begin{figure*}[h!]
	\centering
	\includegraphics[width=1.0\textwidth]{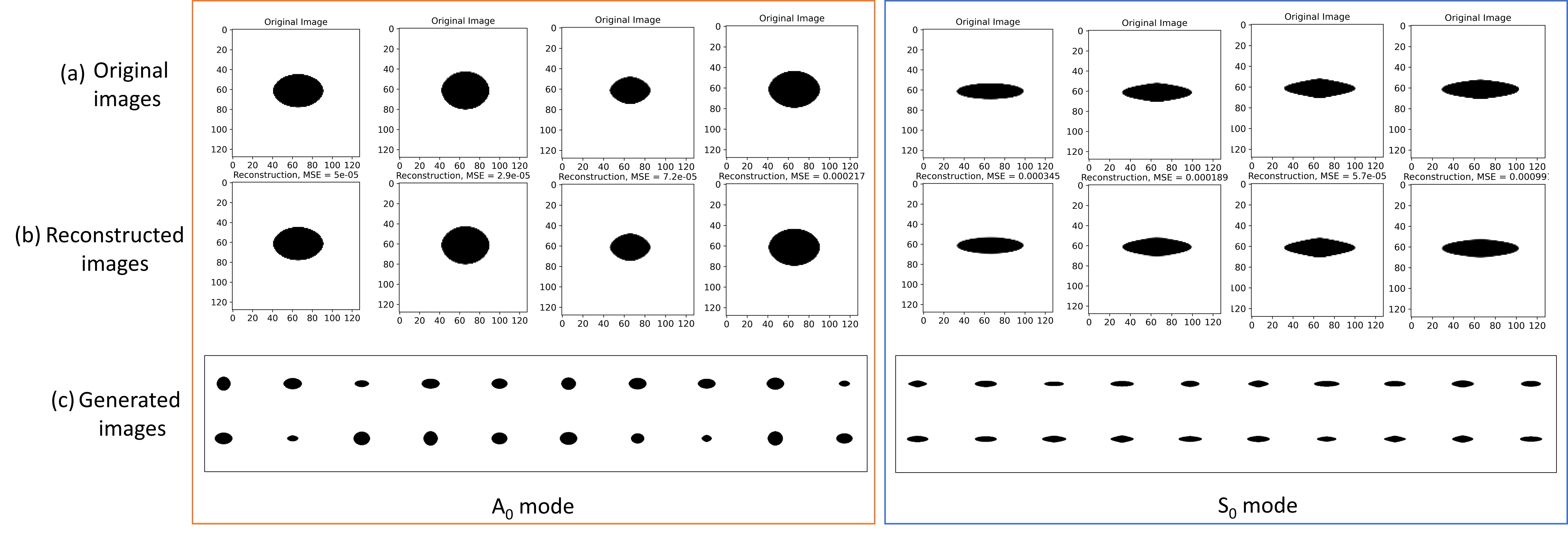}
	\caption{Results from VAE: (a) Original test images (b) reconstructed test images with corresponding MSE (c) New generated images for A0 and S0 modes}
	\label{fig:test}
\end{figure*}

The trained network is tested with dataset-1, containing different polar group velocity representations for commercially available materials. Figs.~\ref{fig:test}(a) and ~\ref{fig:test}(b) shows the test images and their reconstructions from the trained VAE along with the corresponding MSE. It is seen that the network is able to reconstruct the input representations with a low mean squared reconstruction error. 
The reconstructed images are fed into the pretrained inverse prediction model \cite{rautela2021inverse}. The material property predictions corresponding to the original and reconstructed images match very well with less than 0.25\% absolute difference. This verifies the exceptional reconstruction ability of the trained VAE. The source code along with the dataset usage manual is available the the weblink**.

\thispagestyle{github}
%\fancyfoot[L]{center of the footer!}

\begin{figure*}[h!]
	\centering
	\begin{minipage}[b]{0.3\textwidth}
		\centering
		\includegraphics[width=1.0\textwidth]{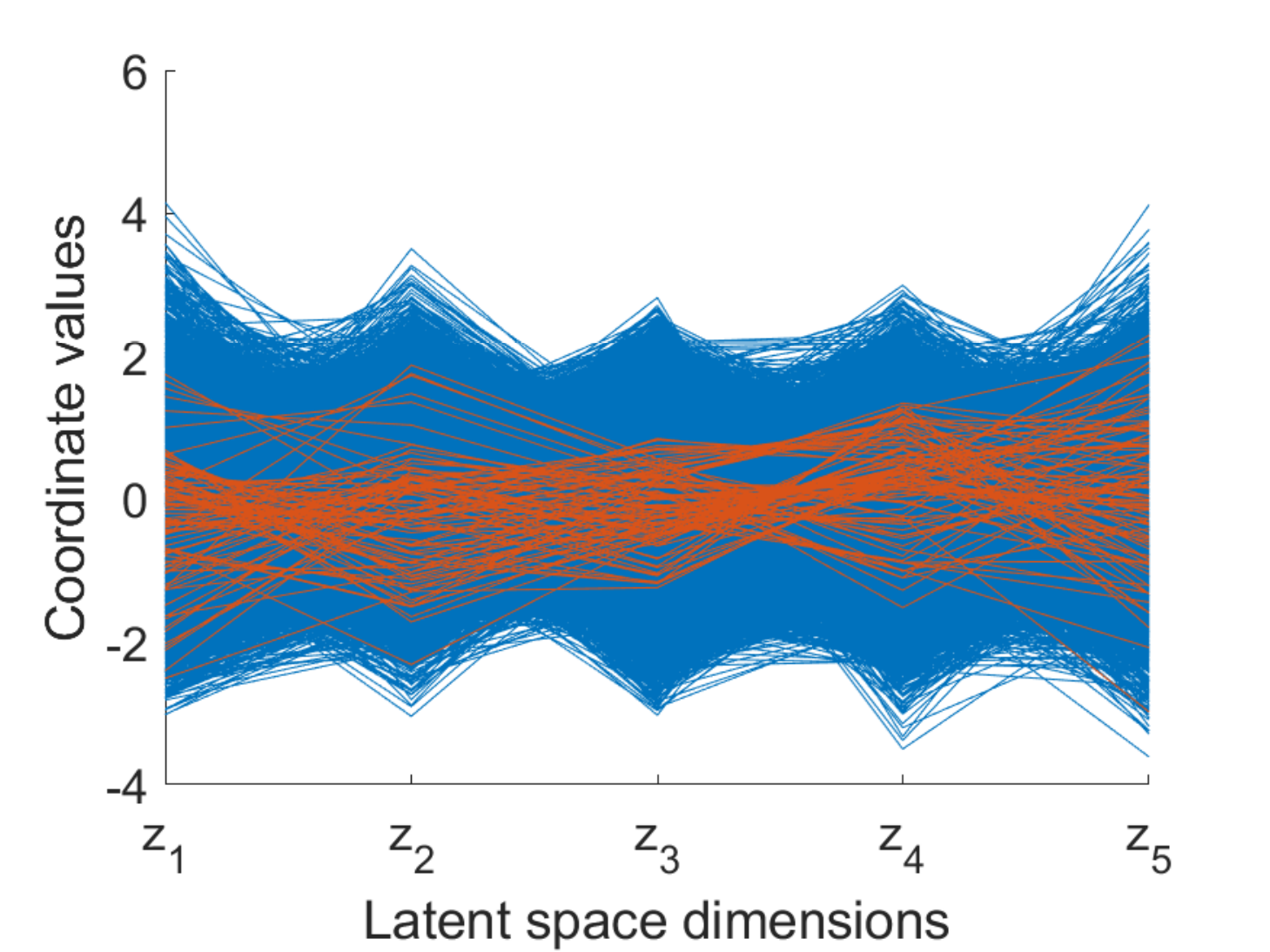}
	\end{minipage}
	\begin{minipage}[b]{0.3\textwidth}
		\centering
		\includegraphics[width=1.0\textwidth]{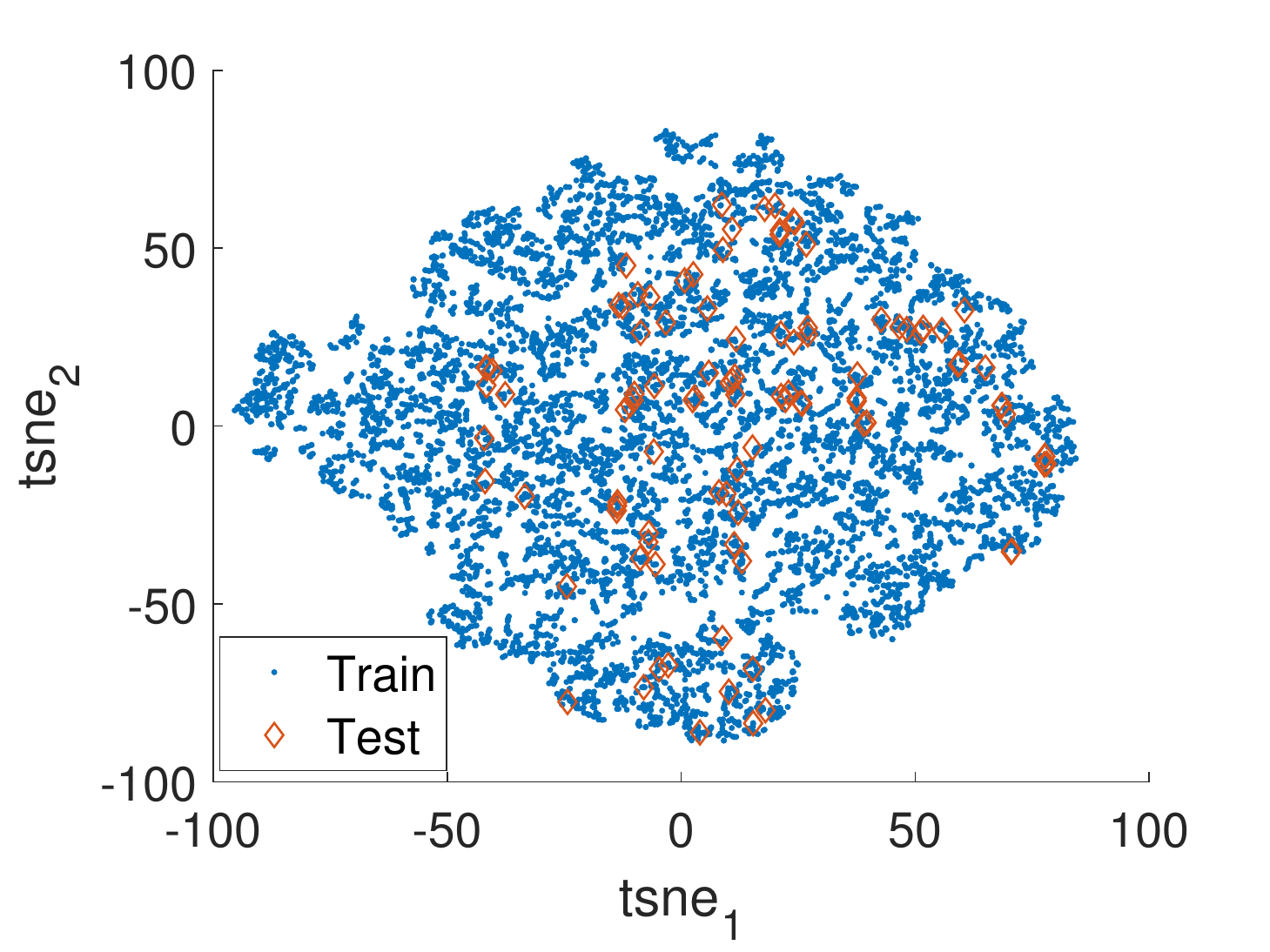}
	\end{minipage}
	\begin{minipage}[b]{0.3\textwidth}
		\centering
		\includegraphics[width=1.0\textwidth]{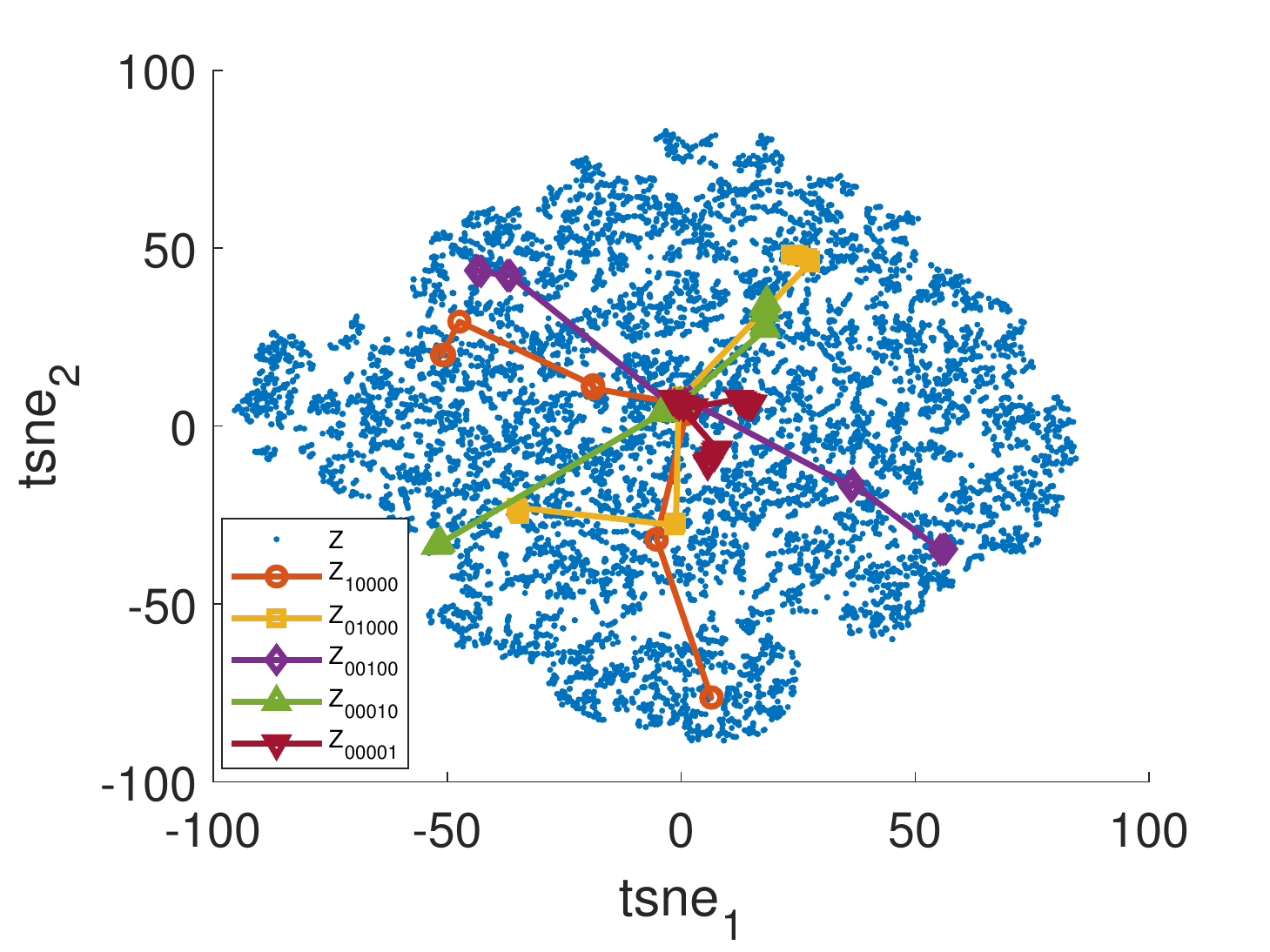}
	\end{minipage}
	\caption{(a) 5D latent space visualization with parallel coordinates plot (b) 2D t-SNE of the latent space (c) t-SNE based visualization of directional sampling}
	\label{fig:latentspace}
\end{figure*}

As discussed earlier, the material properties in dataset-2 are generated randomly from a uniform distribution with bounds set based on the material properties of industrially available materials (dataset-1). Polar group velocity representations are collected from SMM corresponding to the above-mentioned material properties. Therefore, the latent points of the test set should lie well inside the latent space of the training set. Fig.~\ref{fig:latentspace}(a) presents a parallel coordinates plot to visualize the 5-dimensional latent space of the trained VAE. Every line in the plot represents a 5D point in the latent space. Blue and orange lines correspond to the training (dataset-2) and testing set (dataset-1). It is seen from the figure that the latent points of the test set lie well inside the latent space of the training set. To further enhance the visualization, the 5D latent space is transformed to a 2D manifold using t-distributed stochastic neighbor embedding (t-SNE) \cite{van2008visualizing}. Fig.~\ref{fig:latentspace}(b) shows the t-SNE based visualization of the latent space. The test set points lie well inside the boundaries of the training set.

\subsection{Deep generation of guided wave representations}

With a continuous, complete and organized latent space of a well trained VAE, enormous amount (theoretically infinite) of representations can be generated by sampling the latent space. The performance of the generator in generating new polar representations can be evaluated using global Monte Carlo sampling and directional equally-spaced sampling followed by decoding.

\subsubsection{Global Monte Carlo sampling}
Based on Fig.~\ref{fig:latentspace}(a), most of the points resides within [-2 2] bounds for all the five dimensions. In Global Monte Carlo sampling, the latent space is sampled randomly from a uniform distribution between these bounds. The sampled points are fed into the trained decoder to generate new polar representations. Fig.~\ref{fig:test}(c) shows 20 generated images for both A0 and S0 modes. It can also be verified that the generated samples are able to comply with the two-axis symmetry of the polar group velocity representations. The generated images looks realistic, which shows exceptional capabilities of the trained generator. In order to understand the shape and size variations of the generated images in the latent space, directional equally-spaced sampling is performed in different directions.

\subsubsection{Directional equally-spaced sampling}
In directional equally-spaced sampling, the latent space is sampled in equal-spaced steps between the bounds in a particular direction. Using this approach, we can generate multiple representations as well as observe the patterns along particular directions. For instance, $Z_{10000}$ is the direction along with only $z_1$ is changing. Similarly, $Z_{01000}$, $Z_{00100}$, $Z_{00010}$, $Z_{00001}$ are the directions where only $z_2$,$z_3$,$z_4$, and $z_5$ is changing, respectively with no changes in other directions. The movement along these directions can be visualized with t-SNE in Fig.\ref{fig:latentspace}(c). The color-coded lines represent different directions in transformed latent space. One important point to note down is that linear equally-space walk in the latent space along a particular direction does not correspond to a similar movement in the t-SNE-based transformed latent space. This is because t-SNE is a nonlinear dimensionality reduction technique. Therefore, none of the lines representing directions is linear, and the markers representing the steps are not equally spaced.

\begin{figure*}[h!]
	\centering
	\includegraphics[width=1.0\textwidth]{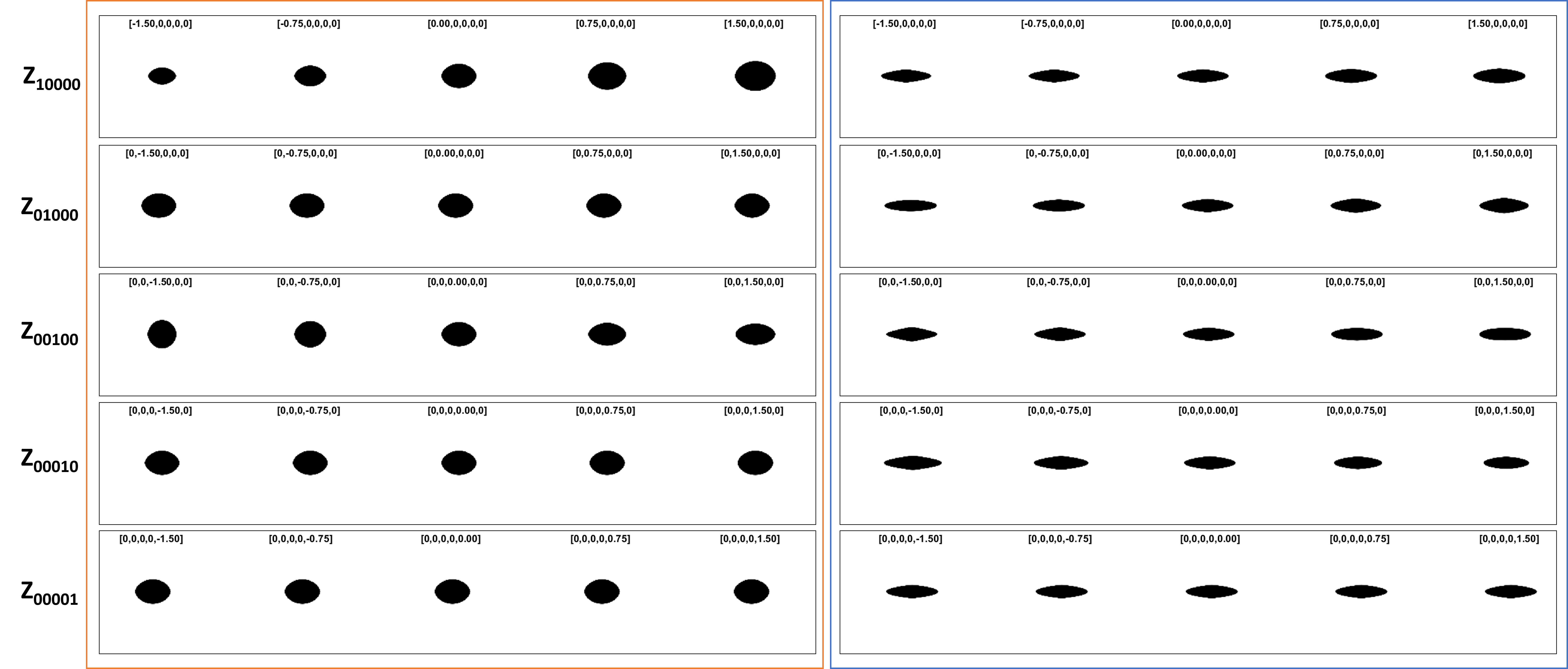}
	\caption{Results from VAE: (a) Original test images (b) reconstructed test images (c) New generated images for A0 and S0 modes}
	\label{fig:directionalImgs}
\end{figure*}

The sampled latent point is fed into the trained decoder to generate the representations along different directions. Fig.~\ref{fig:directionalImgs} shows the generated representations. It can be seen from the figure that the decoder is able to generate realistic polar group velocity representations. 
\begin{figure*}[h!]
	\centering
	\includegraphics[width=0.8\textwidth]{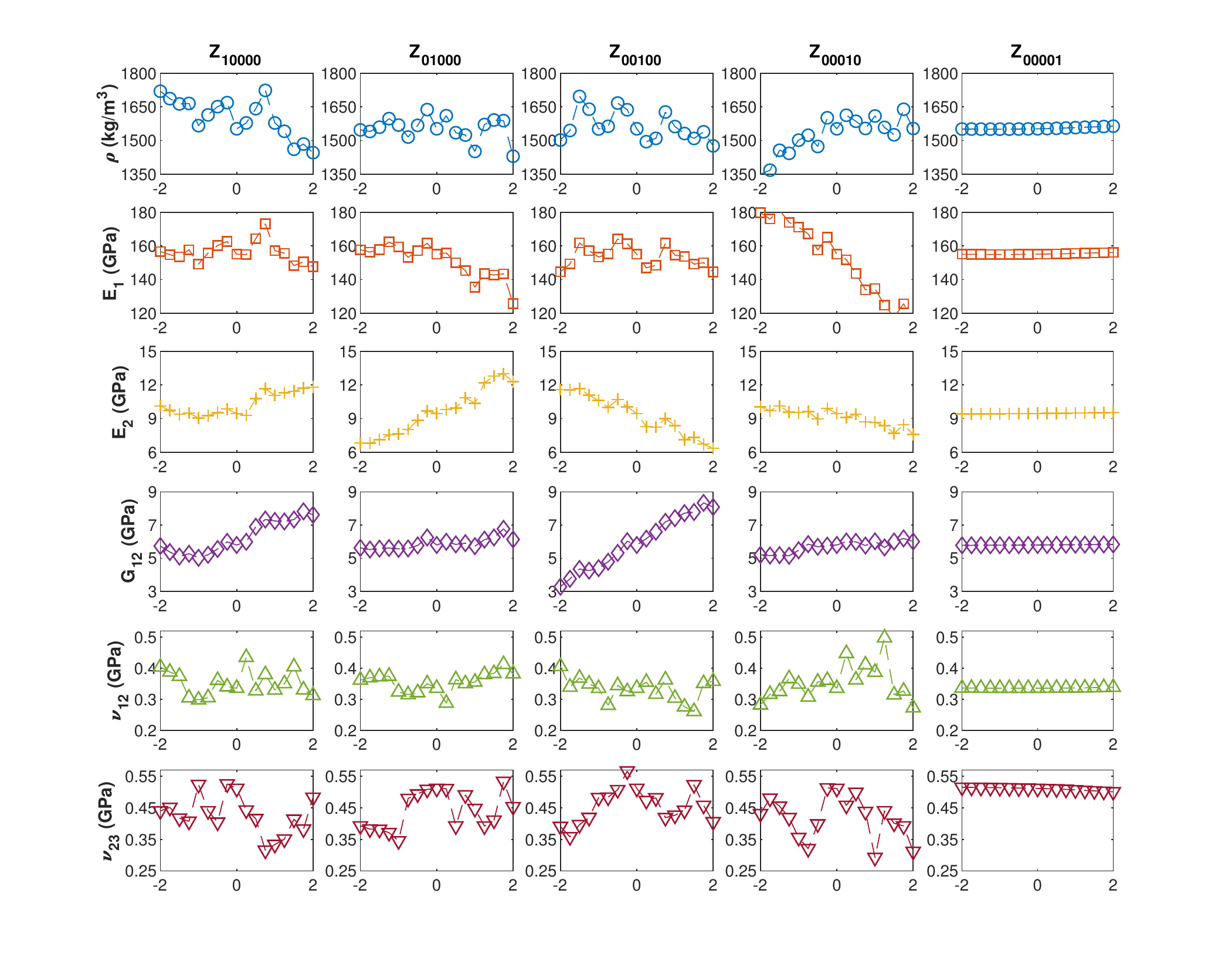}
	\caption{Material properties of the generated composites using directional sampling of the latent space followed by the decoder and inverse property predictor.}
	\label{fig:directionalProp}
\end{figure*}
It is observed from Fig.~\ref{fig:directionalImgs} that there are substantial changes in the generated samples along with different directions. It is seen that there are noticeable shape and size gradual variations in both $A_0$ and $S_0$ representations while moving in different directions. For instance, the size of the $A_0$ is increasing in $Z_{10000}$ direction. Also, the shape of $A_0$ mode is getting flatter in $Z_{01000}$ direction. Similar variations can be seen in the $S_0$ modes. However, in order to quantify these inherent patterns, investigating the material property space is more convenient. The generated samples are fed into the pretrained inverse property predictor \cite{rautela2021inverse}. Fig.~\ref{fig:directionalProp} shows the six different material properties along different directions. The rows and columns of the figure represent material properties and directions, respectively. Some interesting observations can be noted from the figure. It can be seen that the density variation along $Z_{10000}$ has a decreasing trend, whereas, along $Z_{00010}$, it has an increasing trend. Similarly, $E_1$ is decreasing along $Z_{01000}$ and $Z_{00001}$, $E_2$ is increasing along $Z_{01000}$ and decreasing along $Z_{00100}$, shear modulus ($G_{12}$) is increasing along $Z_{10000}$ and $Z_{00100}$. There is no substantial variations seen for Poisson's ratios $\nu_{12}$ and $\nu_{23}$. Also, there are no significant changes in any of the material properties along $Z_{00001}$. The inverse property prediction presented in Fig.~\ref{fig:directionalProp} is satisfactorily under the predefined material bounds used to generate the dataset-2 from SMM. These were [1304 1760] kg/$m^3$, [105 184] GPa, [6 14] GPa, [3 9] GPa, [0.20 0.52] and [0.23 0.59] for $\rho$, $E_1$, $E_2$, $G_{12}$, $\nu_{12}$ and $\nu_{23}$ respectively. 

Designing composite materials is an optimization problem that aims to search for the best combination parameters (such as material and geometric properties) to fulfill the design criteria \cite{jones2018mechanics}. Most of such problems are solved with different optimization schemes where the expensive forward simulations are performed multiple times until the optimum is reached \cite{ghiasi2010optimum}. With our proposed approach, the expensive simulator can be replaced with a deep generator and a deep property predictor. In Ref.~\cite{wang2020deep}, the authors proposed microstructure, graded family, and multiscale system design using variational autoencoder (VAE), property predictor, and a search scheme. In similar directions, the property predictor and the generator can be used along with a search scheme to design lightweight, impact-resistant composite structures. For such design criteria, lower group velocities of $A_0$ and $S_0$ modes along with a lower density of the composite material are some of the most crucial requirements \cite{li2020design,iqbal2020flexural}. Lower group velocities help in steering and delaying the response from the impact force \cite{koo1998vibration,gopalakrishnan2019longitudinal}. With the help of a complete, continuous, and organized low-dimensional latent space, different points can be sampled, followed by decoding to generate group velocity representations in no time. Out of the generated images, those representations with a minimum area (in terms of pixels) can be selected, such that the material also has the minimum density. However, since the scope of this work is limited to generation, material design problems will be investigated as a part of future work.

\section{Conclusion}\label{sec:conclusion}
In this paper, we have proposed a deep generation approach for the generation of new and realistic polar group velocity representations. We have performed physics-based simulations of wave propagation behavior using the stiffness matrix method. The dataset (dataset-2) containing polar group velocity representations corresponding to material properties randomly sampled from a uniform distribution is used to train the variational autoencoder. A dataset (dataset-1) containing representations associated with industrially available composite materials is utilized to test the reconstruction ability and latent space mapping of the trained VAE. Global Monte Carlo and directional equally-space latent space sampling followed by decoding are utilized for the generation of realistic representations. It is seen that the VAE has exceptional generation capabilities for polar group velocity representations. The directional sampling of the latent space reveals inherent patterns hidden in the latent space. The sampling of the latent space in a particular direction reveals shape and size gradual variations of $A_0$ and $S_0$ representations. Different trends are observed in the material properties while moving in particular directions. A detailed investigation of the latent space scanning to understand these patterns will be conducted as a part of future work. The proposed approach can be helpful in accelerating and scaling the design of lightweight impact-resistant structures. Such interesting directions will be studied as a part of upcoming research.

\bibliographystyle{IEEEtran}
\bibliography{mybibfile}

\end{document}